\documentclass[a4paper]{jpconf}
\usepackage{graphicx}

\usepackage{tikz}
\usepackage{graphicx, epstopdf}

\usepackage{xmpmulti}
\usepackage{amsmath, amssymb, amsfonts}
\usepackage{pdfsync}

\newcommand{\half}{\mbox{$\frac{1}{2}$}}

\newcommand{\beq}{\begin{equation}}
\newcommand{\eeq}{\end{equation}}

\newcommand{\const}{{\rm const. \,}}

\newcommand{\R}{\mathbb{R}}

\newcommand{\Z}{\mathbb{Z}}

\begin{document}
\title{Superfluidity and BEC in a Model of Interacting Bosons in a Random Potential 
\footnote{For the Proceedings of 24st INTERNATIONAL LASER PHYSICS WORKSHOP, Shanghai, August 21-25, 2015}}

\author{Martin K\"onenberg$^1$, Thomas Moser$^2$, Robert Seiringer$^3$ and {Jakob Yngvason}$^4$ }

\address{$^1$Department of Mathematics and Statistics, Memorial University St. John's, NL, Canada; present address: Fachbereich  Mathematik,  Universit\"at  Stuttgart,  Germany,\\ $^2$Faculty of Physics,
University of Vienna, Boltzmanngasse 5, 1090 Vienna, Austria; present address: IST Austria, 
Am Campus 1, 3400 Klosterneuburg, Austria,\\ $^3$IST Austria, 
Am Campus 1, 3400 Klosterneuburg, Austria,\\ $^4$Faculty of Physics,
University of Vienna, Boltzmanngasse 5, 1090 Vienna, Austria}

\ead{$^1$martin.koenenberg@mathematik.uni-stuttgart.de, $^2$tmoser@ist.ac.at, $^3$robert.seiringer@ist.ac.at, $^4$jakob.yngvason@univie.ac.at}

\begin{abstract}

We present a mathematically rigorous analysis of the superfluid properties of a Bose-Einstein condensate in the many-body ground state of a one-dimensional model of interacting bosons in a random potential. 
\end{abstract}
\section{Introduction}
The relation between Bose-Einstein condensation (BEC) and superfluidity has been an intriguing question since the discovery of these quantum phenomena in the 1920's and 30's. Although both often occur simultaneously, they are definitely not the same thing: Liquid helium 4 is almost a complete superfluid near absolute zero while the BEC fraction is less than 10 \% \cite{H}. Also, a one-dimensional hard-core Bose gas (Tonks-Girardeau gas) is an example of a complete superfluid where BEC is absent \cite{Le}. On the other hand there is both theoretical and numerical evidence that a random external potential may destroy superfluidity while BEC prevails \cite{ABCG, KT, yukalovgraham, YY}. A mathematically study of this assertion is, however, hampered by the fact that  rigorous proofs of BEC for systems of interacting particles are notoriously difficult and have only been achieved in a few special cases, see in particular \cite{DLS, KLS, ALSSY, LS02}.  Also, rigorous results on the effects of randomness on interacting many-body systems are scarce, contrary to the situation for single particle random 
Schr\"odinger operators where the literature is wast. We refer to \cite{SYZ, KMSY, SS} for lists of representative references on these subjects.

The present contribution summarizes results obtained in \cite{KMSY} (see also \cite{SYZ, SYZ2}) on the effects of a random potential on the superfluid behavior of an interacting Bose gas  in a one-dimensional model. Our main result is that in a certain parameter regime the superfluid fraction can be arbitrarily small while  BEC is complete. In another regime there is both complete BEC and complete superfluidity, despite strong disorder. In the course of the proof we derive a general formula for the superfluid fraction in terms of the wave function of the condensate. Our results are based on this formula combined with energy estimates.

\section{Concepts}

{\it Bose Einstein condensation} in a many-body quantum state means, by definition, macroscopic occupation of some single one-particle state in the limit of large particles numbers  \cite{PO}. Mathematically this is expressed  through the 
{\it one-particle density matrix} of the (pure or mixed) many-body state $\langle\,\cdot\,\rangle$, defined as 
\beq \rho^{(1)}(x,x')=\langle a^\dagger(x) a(x')\rangle=\sum_{i=0}^\infty N_i\, {\psi_i(x)}^*\psi_i(x'),\eeq
 where the eigenvalues $N_0\geq N_1\geq\cdots$, $\sum_{N_i}=N$, are the occupation numbers of the (orthonormal) natural orbitals $\{\psi_i\}$ that are the eigenfunctions of $\rho^{(1)}$.

The {\it condensate fraction} is the relative occupation, $N_0/N$, of the orbital $\psi_0$ with the highest eigenvalue, $N_0$. BEC means that $N_0/N=O(1)$ in the sense that $N_0\geq c N$ with some $c>0$ as $N\to \infty$. The function $\psi_0$  is called the {\it wave function of the condensate}.

It is important to note that the above definition of BEC requires  some specified dependence of the parameters of the Hamiltonian and  the many-body state on $N$ as $N\to\infty$. The standard thermodynamic limit means that the volume increases proportionally to the particle number so that the particle density is kept constant. For experiments with cold atomic gases in traps another limit is relevant, however, and easier to handle but still far from trivial. This is the Gross-Pitaveskii (GP) limit \cite{LSSY}, which can be regarded as a combination of a thermodynamic and a weak coupling limit. In this limit, which is the one considered in the present contribution, the ground state energy per particle is comparable to the energy gap of the free Hamiltonian in the trap.

Concerning the definition of {\it superfluidity} we note that this concept can mean  different things \cite{Leggett}:

\begin{itemize}
\item Flow at a finite velocity without friction (non-equilibrium phenomenon)
\item Non-classical response to an infinitesimal boost or rotation (equilibrium phenomenon)
\end{itemize}
Here we consider  superfluidity only in the sense of the second definition. 
The {\it superfluid mass density} $\rho_{\rm s}$ (at rest) is then defined through the response of the free energy (or, at $T=0$, the  ground state energy) to a small boost $v$:
\beq F(v)=F(0)+\half \hbox{$mN$}\,(\rho_{\rm s}/\rho) \hbox{$v^2$}+o(v^2).\eeq
The boost is mathematically implemented by the substitution
\beq p_i\rightarrow p_i-mv\eeq
in the Hamiltonian, assuming {\it periodic boundary conditions} in the direction $\vec e$ of 
the boost with period $\Lambda$, say.
Experimentally, the boost can conveniently be realized in a thin tube bent to a circular container that is brought into slow rotation. The superfluid fraction can be derived from the moment of inertia of the system.

Equivalently one can consider the original Hamiltonian in a box (interval) but with  {\it twisted boundary conditions}:
\beq \Psi(\dots, x_{i-1}, \Lambda\vec e, x_{i},\dots)=e^{-\mathrm i\varphi}\Psi(\dots, x_{i-1}, 0, x_{i},\dots)\eeq
with $\varphi=mv \Lambda/\hbar.$

In the following we discuss a simple model of interacting particles where the effect of a random potential on BEC and superfluidity can be investigated mathematically. 

\section{The Model}

As in \cite{SYZ, KMSY} the model is a one-dimensional gas of bosons with contact interaction (Lieb-Liniger model \cite{LL}) on the unit interval but with an additional external random potential $V_\omega$. The Hamiltonian on the Hilbert space $L^2([0,1], dz)^{\otimes_{\rm symm}^N}$ is
\beq\label{ham}  H=\sum_{i=1}^N\left(-\partial_{z_i}^2+V_\omega(z_i)\right)+\frac{{\gamma}} N\sum_{i<j}\delta(z_i-z_j)\eeq
 with a coupling constant $\gamma\geq 0$ and  {\it periodic} boundary conditions. This can also be regarded as a  model  of a gas on a ring (or a thin tube bent to a circle) of radius $1/2\pi$.  Units have been chosen so that $\hbar=1$ and the mass is $m=\half$.

The random potential will be taken to be
\beq V_\omega(z)=\sigma\sum\delta(z-z_j^\omega)\eeq with $\sigma \geq 0$ independent of the random sample $\omega$ while the point obstacles $\{z_j^\omega\}$ are Poisson distributed with density ${\nu}\gg 1$.  
i.e., their mean distance is $\nu^{-1}\ll 1$. 

Besides $N$, the model has three parameters:

\begin{itemize}
\item$\gamma$: Strength of the interaction

\item$\nu$: Density of obstacles

\item$\sigma$: Strength of the random potential
\end{itemize}
In the $N\to\infty$ limit the ground state energy and the wave function of the condensate is described by a  
Gross-Pitaevskii (GP) energy functional \eqref{gp} as discussed further below. 

In \cite{SYZ} it was proved that the  ground state energy becomes deterministic, i.e.\ almost surely independent of $\omega$,  if the parameters satisfy 
\begin{equation}\label{standard}
    \nu\gg 1\,, \quad \gamma\gg \frac{\nu}{\left(\ln \nu\right)^2}\,, \quad 
    \sigma \gg \frac{ \nu}{1+ \ln \left( 1+ \nu^2/\gamma\right) }\,.
  \end{equation}
  We shall refer to these conditions as the {\it standard conditions} and assume them throughout.


\section{The Main Results}
Our main results about the model \eqref{ham} are as follows:
{\it \begin{itemize}
\item[1.] In the whole parameter range there is complete BEC in the ground state.
\item[2.] If $\gamma\lesssim \nu^2$ the superfluid fraction is arbitrarily small, i.e., it goes to zero under the standard conditions \eqref{standard}.
\item[3.] The same holds for $ \nu^2\ll \gamma\ll \nu^4$ provided $\sigma\gg(\gamma/\nu^2)^2\gamma^{1/2}$.
\item[4.] If $\gamma\gg (\sigma \nu)^2$ there is complete superfluidity, i.e., the superfluid fraction tends to 1.
\end{itemize}}
These results are illustrated in Fig.\ 1. It should be noted that our results describe asymptotic properties for large values of the parameters.  The boundaries of the red and green areas in the figure should therefore not be interpreted as sharp phase boundaries.


\begin{figure}[ht]
\center
\fbox{\includegraphics[width=7.5cm]{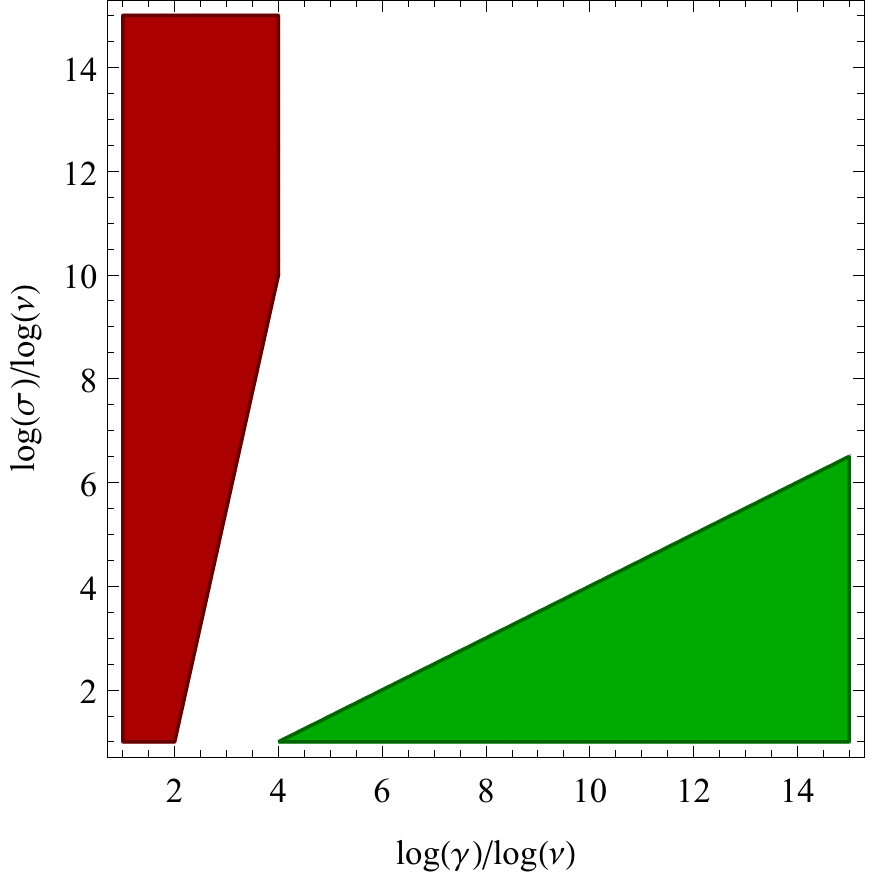}}
\caption{Red: Absence of superfluidity. Green: Complete superfluidity.}
\label{}
\end{figure}

The general heuristic picture is that 
\begin{itemize}
\item Strong repulsive interaction between the particles (large $\gamma$) tends to make the density uniform and favors superfluidity.
\item Strong randomness (large $\nu$ and $\sigma$) leads to fragmentation of the density that is unfavorable for superfluidity.
\end{itemize}

This heuristics has been  confirmed in \cite{KMSY} by rigorous mathematical analysis. It is notable that
\begin{itemize}
\item BEC survives the fragmentation of the density, i.e., long range correlations prevail although superfluidity may be strongly suppressed.
\item Superfludity may be suppressed even in cases when the density is close to being uniform on the average and thus not ``visibly" fragmented (case 3 above).
\end{itemize}
We note also that complete absence of both BEC and superfluidity has recently been proved for the Tonks-Girardeu gas, i.e.\  for $\gamma/N=\infty$, in a generic random potential \cite{SS}.

\section{Bose-Einstein Condensation}

We summarize here the basic findings of \cite{SYZ} concerning Bose-Einstein condensation  in the ground state of the many-body Hamiltonian \eqref{ham}.

BEC is proved to hold in the GP limit, where $N\to\infty$ and $\gamma$ is fixed, or does not grow too fast with $N$. This applies to an arbitrary positive external potential $V$. 
 
The wave function of the condensate  is the (unique) nonnegative minimizer $\psi_0$ of the Gross-Pitaevskii (GP) energy functional
\beq\label{gp}
\mathcal{E}^{\rm GP}[\psi] = \int_0^1 \left( |\psi'(z)|^2 + V(z) |\psi(z)|^2 + \frac \gamma 2 |\psi(z)|^4 \right) {\mathrm d} z
\eeq
with the normalization $\int_0^1|\psi|^2=1$. 
The minimizer $\psi_0$ is also the ground state of the mean field one-particle Hamiltonan
{\beq\label{meanfieldham}
h=-\partial_{z}^2+V(z)+\gamma |\psi_0|^2-\frac\gamma 2\int_0^1 |\psi_0|^4
\eeq}
with eigenvalue $e_0=\mathcal{E}^{\rm GP}[\psi_0]$.


The average occupation of the one-particle state $\psi_0$ in the many-body ground state $\Psi_0$ is (by definition) 
\beq N_0=\langle \Psi_0, a^\dagger(\psi_0)a(\psi_0)\Psi_0\rangle.\eeq
Bose-Einstein condensation in the GP limit follows from an estimate of the depletion of the condensate that was derived in \cite{SYZ}:
\beq \left(1-\frac {N_0}N\right)\leq ({\rm const.}) \frac {e_0}{e_1-e_0}
N^{-1/3}\min\{\gamma^{1/2},\gamma\}
\eeq
where $e_1$ is the second lowest eigenvalue of the mean field Hamiltonian $h$. Moreover,  the ground state energy of $H$ per particle   converges to the GP energy $e_0$.


\section{Superfluidity}

With an imposed velocity field $v$ (moving walls) the Hamiltonian becomes:
\beq\label{hamv}
H_v=\sum_{j=1}^N\left\{({\mathrm i}\partial_{z_j}
+v)^2+V(z_j)\right\}+\frac \gamma N\sum_{i<j}\delta(z_i-z_j)
\eeq
on $L^2([0,1],{\mathrm d}z)^{\otimes_{\rm symm}^N}$ with periodic boundary conditions.
We denote by  $E_0^{\rm QM}(v)$ its ground state energy and by
$e_0(v)$  the corresponding  energy of the modified GP functional
\beq\label{modGP}
\mathcal{E}^{\rm GP}_v[\psi] = \int_0^1 \left( |{\mathrm i}\psi'(z)+ v\psi(z)|^2 + V(z) |{}{\psi}(z)|^2 + \frac \gamma 2 |\psi(z)|^4 \right) {\mathrm d} z.
\eeq


For small enough $v$, $\mathcal{E}_v^{\rm GP}$ has a unique minimizer, denoted by $\psi_v$, and $e_0(v)$ is equal to the ground state energy of the boosted mean field Hamiltonian
\beq
h_v=({\mathrm i}\partial_{z}+v)^2+V(z)+\gamma |\psi_v(z)|^2-\frac\gamma 2\int_0^1 |\psi_v|^4.
\eeq
Using the diamagnetic inequality, $ |({\mathrm i}\partial_{z}+v)\psi|^2\geq |\partial_z|\psi||^2$, 
one shows in an analogous way to the $v=0$ case:
\beq\label{lowerbd}
E_0^{\rm QM}(v)/N\geq e_0(v)(1-(\const)N^{-1/3}\min\{\gamma^{1/2},\gamma\}).
\eeq
We conclude that in the GP limit the superfluid fraction at $v=0$
{\beq\label{sfracQM}
\rho_{\rm s}/\rho=\lim_{v\to 0}\frac 1{v^2}\lim_{N\to\infty}\frac 1N(E_0^{\rm QM}(v)-E_0^{\rm QM}(0))
\eeq}
is the same as the corresponding quantity derived from the GP energy, i.e., 
\beq\label{sfracgp}
\rho_{\rm s}/\rho=\lim_{v\to 0}\frac 1{v^2}(e_0(v)-e_0(0)).
\eeq


\section{A closed formula for the superfluid fraction}

We claim that
\begin{equation}\label{defsf}
\rho_{\rm s}/\rho = \left( \int_0^1 |\psi_0(z)|^{-2} dz \right)^{-1}
\end{equation}
This provides and explicit connection between the wave function of the condensate and the superfluid fraction  at $v=0$. The proof is short enough to be reproduced here in full:

We start with the variational equation for $\psi_v$:
\beq
(i \partial_z + v)^2 \psi_v(z) + V(z) \psi_v(z) + \gamma |\psi_v(z)|^2 \psi_v(z) = \mu \psi_v(z).
\eeq
Multiplying  by the conplex conjugate $\psi^*_v$ and taking the imaginary part gives
\beq
\partial_z \left( v |\psi_v(z)|^2 - \Im [ \psi^*_v(z) d\psi_v(z)/dz] \right) = 0.
\eeq
Hence there exists a constant $C\in \R$ such that 
\beq
 \Im [ \psi^*_v(z) d\psi_v(z)/dz] = v |\psi_v(z)|^2 - C.
 \eeq
 

Since 
\begin{equation}\label{primeE}
de_0(v)/dv = 2 v - 2  \int_0^1 \Im [\bar \psi_v(z) d\psi_v(z)/dz] dz
\end{equation}
 we actually see that $C=\half de_0(v)/dv$.
  
Now $\psi_v$ has no zeroes for small $v$ so we can divide by $|\psi_v(z)|^2$ and obtain
 \beq
 S'(z): = \frac{  \Im [ \psi^*_v(z) d\psi_v(z)/dz]  }{ |\psi_v(z)|^2}  = v - \frac C {|\psi_v(z)|^2}.
 \eeq
 Since $S'$ is, in fact, the derivative of the phase of $\psi_v$ we have, due to the periodic boundary conditions, 
 \beq
 \int_0^1 S'(z) dz = 2\pi n 
 \eeq
with $n\in \Z$, and in fact $n=0$ for small enough $v$. 
Therefore
 \beq
 v = C \int_0^1 |\psi_v(z)|^{-2} dz. 
 \eeq
 
This gives
\beq
e_0'(v) = 2 C = 2 v \left(  \int_0^1 |\psi_v(z)|^{-2} dz  \right)^{-1}
\eeq
and thus 
\beq\label{27}
\rho_{\rm s}/\rho = \lim_{v\to 0} \frac{ e_0'(v)}{2v}= \left( \int_0^1 |\psi_0(z)|^{-2} dz \right)^{-1}.
\eeq\hfill $\Box$

\section{Complete superfluidity for $\gamma\gg (\sigma\nu)^2$}

The formula \eqref{27} leads to a proof of complete superfluidity for $\gamma\gg (\sigma\nu)^2$ in the following way. First, taking 1 as a trial function for the GP functional  \eqref{modGP} we see that the GP energy is bounded from above by $\int V_\omega+\frac \gamma 2$.
Using this bound and the fact that $\psi_0$ is a GP minimizer a simple calculation leads to
\beq\label{sobolev}
 \frac { \| |\psi_0|^2 - 1 \|_\infty^2 } { \sqrt{ 1 + \|\, |\psi_0|^2 - 1 \|_\infty}} \leq \frac {2^{3/2}} {\sqrt{\gamma}} \int_0^1 V_\omega\sim (\sigma\nu)/\gamma^{1/2}
\eeq
where $\|\, \cdot \|_\infty$ is the sup norm and the estimate of $\int V_\omega$ holds with high probability.
Hence we see that
the superfluid fraction tends to 1 in probability if $$\gamma\gg (\sigma\nu)^2.$$

In general, however, it turns out not to be sufficient that $\gamma\gg \nu^2$, 
although in that case it can be shown that $\| |\psi_0|^2 - 1 \|_1\to 0$.


\section{Absence of superfluidity}

If $\mathcal I$ is any (measurable) subset of $[0,1]$ with length $|\mathcal I|$ we have
$$|\mathcal I|^2=\left(\int_{\mathcal I}|\psi_0||\psi_0|^{-1}\right)^2\leq \int_{\mathcal I}|\psi_0(z)|^2\cdot \int_{\mathcal I}|\psi_0(z)|^{-2}$$
and hence, by \eqref{27},
\beq\label{superfbound} \rho_{\rm s}/\rho\leq \frac{\int_{\mathcal I}|\psi_0(z)|^2 dz}{|\mathcal I|^2}.
\eeq

To prove that superfluidity is {small} we have therefore to identify subsets $\mathcal I$ such that $\int_{\mathcal I}|\psi_0(z)|^2 dz$ is {small}, while $|\mathcal I|$ is not too small. 

The random points $z_j^\omega$ split the interval $[0,1]$ into subintervals
$\mathcal I_j=[z_j^\omega, z_{j+1}^\omega]$ of various lengths $\ell_j=z_{i+1}^\omega-z_{j}^\omega$ that  are i.i.d. random variables with probability distribution
{\beq dP_\nu(\ell)=\nu e^{-\nu\ell}\, d\ell.
\eeq}
We take
{\beq \mathcal I=\bigcup_{j:\ell_j\leq\tilde \ell}\mathcal I_j
\eeq}
with a {suitably chosen} $\tilde\ell$. 
The average length of $\mathcal I$ (which is a random variable because the points $z_j^\omega$ are random) is
\beq \label{3.12} L=\nu\int_0^{\tilde\ell}\ell dP_\nu(\ell)=1-(1+(\nu\tilde\ell))e^{-\nu\tilde\ell}).\eeq
In particular it tends to 1 if and only if $\tilde\ell\gg \nu^{-1}$.


With the notation
{\beq\label{gpmass}
 n_j^{\rm GP}=\int_{\mathcal I_j}|\psi_0(z)|^2dz\eeq}
we define
\beq \label{3.5} N_{\rm small}=\int_{\mathcal I}|\psi_0(z)|^2 dz=\sum_{\ell_j\leq \tilde \ell} n^{\rm GP}_j.\eeq
Note that $\psi_0$ and $n_j^{\rm GP}$ also depend on the random sample $\omega$ for the Poisson distribution of the obstacles but we have suppressed this in the notation for simplicity.

Our estimate on $N_{\rm small}$ is based on estimates on the GP energy. 
The superfluid fraction is then bounded from above by  $N_{\rm small}/L^2$.

\subsection{The case $\gamma\lesssim \nu^2$}
One chooses
{\beq\label{3.2} \tilde\ell=s/\nu
\eeq}
with a suitable $s>0$ and proceeds in the following steps:
\begin{itemize}
\item Split the GP energy into contributions from \lq large' intervals of length $\geq\tilde \ell$ and \lq small' intervals of length 
$<\tilde \ell$.
\item Estimate each contribution using estimates on an auxiliary GP functional for the density between two obstacles.
\item Using that ${e_0(\gamma,\nu,\sigma)}/\nu^2$ stays bounded  under the standard conditions \eqref{standard} one obtains a bound $o(1)$ on the {mass} $N_{\rm small}$.
\item The total {length} $L$ of the small intervals and its fluctuation are estimated to be $O(1)$ and $o(1)$ respectively.\end{itemize}
The upshot is that
the superfluid fraction ($\leq\,N_{\rm small}/L^2$) tends to 0 in probability if $\gamma\lesssim \nu^2$.


\subsection{The case $\gamma\gg \nu^2$}

Here both the mass and the length of the small intervals are $o(1)$, but one can prove that $N_{\rm small}/L^2\ll 1$ still holds if the obstacle barriers are high enough, namely for
\beq\label{36}\sigma\gg(\gamma/\nu)^4\gamma^{1/2}.\eeq
In addition one needs to know that the {fluctuations} of $L$ are much smaller than the average value and this can be shown to hold for $\gamma\ll \nu^4$. Altogether one obtains:
 If  $\nu^2\ll\gamma\ll \nu^4$ the superfluid fraction tends to 0 in probability provided \eqref{36} holds.


\section{Summary}

We have studied superfluidity in the ground state of a one-dimensional model of bosons with a repulsive contact interaction and in a random potential generated by Poisson distributed point obstacles. 

In the Gross-Pitaevskii (GP) limit this model always shows complete BEC, but depending on the parameters, superfluidity may or may not occur. In the course of the analysis we derived the closed formula \eqref{defsf} for the superfluid fraction, expressed in terms of the GP wave function.

An advantage of the model considered is its amenability to a rigorous mathematical analysis leading to unambiguous statements.  It has its limitations, however: Nothing is said about positive temperatures and the proof of BEC applies to the Gross-Pitaevskii limit rather than the thermodynamic limit. 

Nevertheless, to our knowledge this is the only model where a Bose glass phase in the sense of  \cite{yukalovgraham}, i.e., complete BEC but absence of superfluidity, has been rigorously established so far.

\ack

Support from the  Natural Sciences and Engineering Research Council of 
Canada NSERC (MK, and RS under project P 27533-N27) and from the Austrian Science Fund FWF (JY, under project P 22929-N16) is gratefully acknowledged. 


\end{document}